\documentclass[preprint2]{aastex}

\usepackage{lscape}
\usepackage{multirow}
\usepackage{color}
\usepackage{rotating}
\usepackage{lscape}
\definecolor{gray}{gray}{0.6}

\shorttitle{The ${}^8{\rm Li}+{}^4{\rm He}\to{}^{11}{\rm B}+n$
reaction at astrophysical energies}
\shortauthors{La Cognata et al.}

\begin{document}


\title{Solving the large discrepancy between inclusive and
exclusive measurements of the ${}^8{\rm Li}+{}^4{\rm
He}\to{}^{11}{\rm B}+n$ reaction cross section at astrophysical
energies}


\author{M.La Cognata$^{a,b,c}$,
A.Del Zoppo$^a$,
R.Alba$^a$,
S.Cherubini$^{a,b}$,
N.Colonna$^f$,
A.Di Pietro$^a$,
P.Figuera$^a$,
M.Gulino$^{a,b}$,
L.Lamia$^a$,
A.Musumarra$^{a,b}$,
M.G.Pellegriti$^{a,e}$,
R.G.Pizzone$^a$,
C.Rolfs$^d$
\altaffilmark{1},
S.Romano$^{a,b}$,
C.Spitaleri$^{a,b}$ \and
A.Tumino$^{a,g}$}
\affil{
$^a$INFN-Laboratori Nazionali del Sud, Via S.Sofia 62, I95123 Catania, Italy\\
$^b$Dipartimento di Metodologie Fisiche e Chimiche per l'Ingegneria,
Universit\`a di Catania, I95123 Catania, Italy\\
$^c$Centro Siciliano di Fisica Nucleare e Struttura della Materia, Catania, Italy\\
$^d$Institut fur Physik mit Ionenstrahlen, Ruhr-Universit\"at Bochum, Bochum, Germany\\
$^e$Dipartimento di Fisica e Astronomia, Universit\`a di Catania, I95123 Catania, Italy\\
$^f$INFN-Sezione di Bari, Via Orabona 4, I70126 Bari, Italy\\
$^g$Universit\`a Kore, Enna, Italy
}
\email{DelZoppo@lns.infn.it}

\altaffiltext{1}{Supported in part by Deutsche Forschungsgemeinschaft.}


\begin{abstract}
A solution of the large discrepancy existing between inclusive and exclusive
measurements of the ${}^8{\rm Li}+{}^4{\rm He}\to{}^{11}{\rm B}+n$ reaction cross
section at $E_{cm} <3$ MeV is evaluated. This problem has profound astrophysical
relevance for this reaction is of great interest in Big-Bang and
r-process nucleosynthesis. By means of a novel technique, a comprehensive study
of  all existing ${}^8{\rm Li}+{}^4{\rm He}\to{}^{11}{\rm B}+n$ cross
section data is carried out, setting up a consistent picture in which all the
inclusive measurements provide the reliable value of the cross
section.  New unambiguous signatures of the strong branch
pattern non-uniformities, near the threshold of higher
${}^{11}{\rm B}$ excited levels, are presented and their
possible origin, in terms of the cluster structure
of the involved excited states of ${}^{11}{\rm B}$ and ${}^{12}{\rm B}$
nuclei, is discussed.
\end{abstract}





\begin{figure}
\epsscale{1}
\plotone{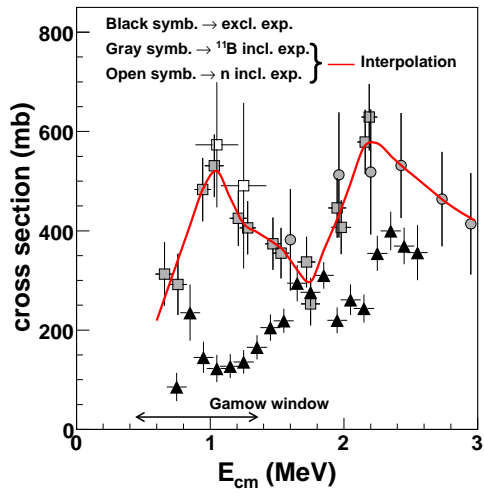}
\caption{\label{fig1}{${}^8{\rm Li}+{}^4{\rm He}\to{}^{11}{\rm B}+n$
reaction cross section data versus $E_{cm}$: $\square$ \citep{LAC08},
\textcolor{gray}{\Large{$\bullet$}} \citep{BOY92}, \textcolor{gray}{ $\blacksquare$}
\citep{GU95}, $\blacktriangle$ \citep{ISH06}. The curve depicts
the interpolated inclusive cross section
$\sigma_{incl}$. The Gamow energy region at $T=2
\times 10^9$ K is shown.}}
\end{figure}

Recently, the total disagreement between inclusive
measurements of the ${}^8$Li+${}^4$He$\to{}^{11}$B+$n$ reaction cross section and exclusive
ones at kinetic energy in the centre-of-mass system $E_{cm} <2$ MeV has been pointed out \citep{LAC08}.
The case is  illustrated in Fig.\ref{fig1}.
The complementary neutron inclusive \citep{LAC08} and $^{11}$B
inclusive \citep{BOY92,GU95} measurements give comparable values of
the reaction cross section. Instead, the exclusive
approach \citep{ISH06}, where  $^{11}$B-n coincidences were
measured, gives cross-section values which, with the exception
of a narrow region around  $E_{cm}\sim 1.7$ MeV, are generally
smaller by a factor $\leq 4$ with respect to the inclusive ones.

The largest discrepancy among the data in Fig.\ref{fig1} is
observed right at energies of astrophysical interest $E_{cm}\cong 1$ MeV.
The measurements at such energies explore the Gamow window of the
${}^8$Li+${}^4$He$\to{}^{11}$B+$n$ reaction at temperatures
$T =(1\div 3)\times 10^9$ K, of great interest in primordial
as well as in other relevant nucleosynthesis sites, in particular
core-collapse supernovae and neutron-star mergers. Within the frame of the
inhomogeneous Big Bang model, still
representing a viable possibility for the early universe \citep{Lara06,MAL88,KAJ90,RAU07},
this reaction could have allowed to overcome the \textit{A}=8 mass gap, thus
providing a possible explanation for the experimental observation of a
non-negligible abundance of heavy elements in the oldest
astrophysical objects (\citet{MAT05} and Refs. therein).
The  magnitude of the cross section is the key information to check the
reliability of these predictions.
The ${}^8$Li+${}^4$He$\to{}^{11}$B+$n$ reaction also plays an
important role in the context of the r-process nucleosynthesis \citep{Sasa06,TER01}.
Currently, the most popular scenario
is neutrino-driven winds from Type II SNe. Anyway,
possibility remains that it could be associated with neutron-star
mergers or gamma-ray bursts, in which the required neutron-rich
conditions can also be realized. Therefore, it is of critical
importance to constrain the parameter space for the r-process to
restrict possible environments. \citet{Sasa06} found that the
${}^8$Li+${}^4$He$\to{}^{11}$B+$n$ reaction leads to a
more efficient production of seed nuclei, so that a larger neutron/seed
ratio is required for a successful r-process.
This, in turn, allows to constrain the entropy per baryon and
the astrophysical site for production of r-process nuclei.

In this Letter, we conclude that inclusive
measurements provide the most reliable estimate of the cross section.

\begin{figure}
\epsscale{1}
\plotone{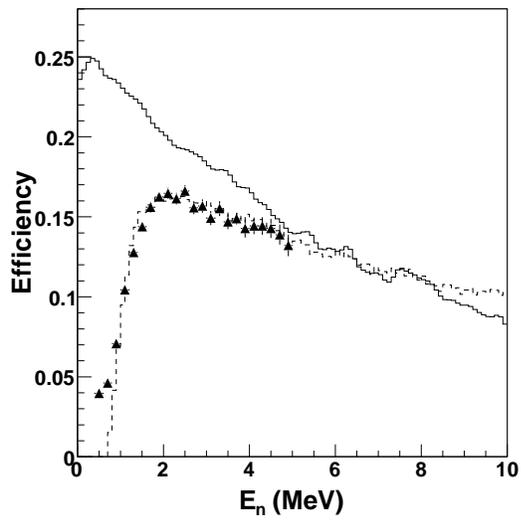}
\caption{\label{fig2} The dependence of the total detection efficiency of the
thermalization counter in \citet{LAC08} (solid histogram) and of the plastic
scintillator arrays in \citet{ISH06,Ash06} (closed triangles) on the laboratory
neutron energy. The dashed histogram is the Montecarlo simulation which better
reproduces the efficiency data \citep{Ash06}.}
\end{figure}

\begin{figure}
\epsscale{1}
\plotone{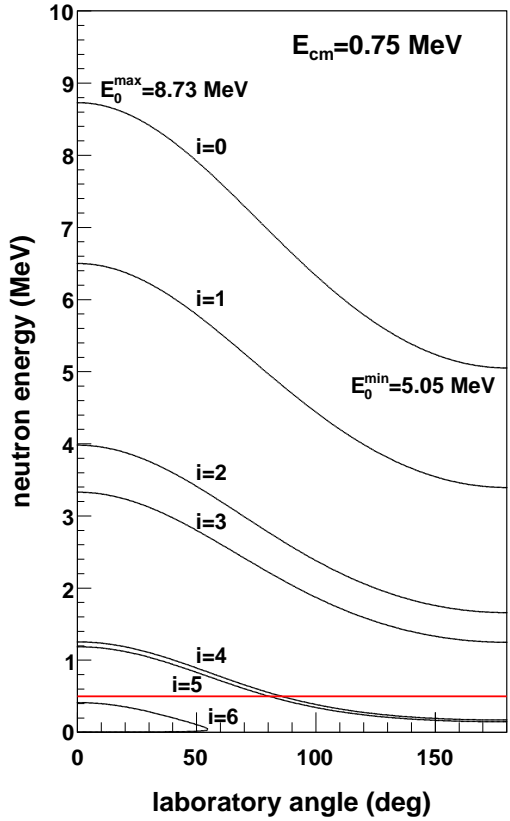}
\caption{\label{fig1bis} Kinematical plot showing the neutron energy-angle correlation
for each branch of the ${}^8{\rm Li}+{}^4{\rm He}\to{}^{11}{\rm B}+n$ reaction at
$E_{cm} = 0.75$ MeV. The minimum and maximum neutron energies allowed by kinematics
in the whole angular range are given for the $i=0$ branch ($^{11}$B ground state).
The red line represent the 0.5-MeV seeming cut on the neutron energy from Fig.\ref{fig2}.
}
\end{figure}

For the ${}^8$Li+${}^4$He$\to{}^{11}$B+$n$ reaction involves the unstable
$^{8}$Li nucleus ($T_{1/2}=840$ ms), the scope of measurements
with present-day facilities is the cross section $\sigma (E_{cm})$
summed over all $^{11}$B+n$_i$ branches,
\begin{equation}
\label{eq0}
\sigma (E_{cm}) =\sum_{i=0}^{i_{max}} \frac{{\rm N}_i^{det}}{\epsilon_i {\rm N}^{tar} {\rm N}^{proj}} \, ,
\end{equation}
$i_{\max}$ denoting the energetically open highest branch at a given $E_{cm}$ ($i=0$
being the branch leading to $^{11}$B ground state).
The involved reaction
branches, identified by the $^{11}$B level sequence in \citet{ISH06}, are
listed in Tab.\ref{tab1}.
Examining possible error sources that might spoil
cross section measurements, the number of detected particles N$_i^{det}$, of
impinging projectiles N$^{proj}$, of target nuclei per
unit surface N$^{tar}$, and the detection efficiencies $\epsilon_i$ contribute to the
overall uncertainty.  Background might illusorily enhance the reaction yields N$_i^{det}$
of inclusive measurements \citep{BOY92,GU95,LAC08}. However, they were performed by
measuring completely different ejectiles ($^{11}$B or neutron),
thus different sources of background are expected. Therefore, the agreement
between them make us confident that the possible background contribution stays
below the uncertainty ranges in Fig.\ref{fig1}.
We also underscore that each of inclusive \citep{BOY92,GU95}
and exclusive \citep{ISH06} data sets has been measured in a single
irradiation, thus any error in their evaluation attributable to N$^{tar}$
and/or  N$^{proj}$ only could lead to a rigid vertical displacement of one
excitation function with respect to the others, excluded by Fig.\ref{fig1}.
Indeed, concordant inclusive \citep{BOY92,GU95} and exclusive \citep{ISH06} cross section
measurements at E$_{c.m.}\sim 1.7$ MeV (see Fig.\ref{fig1})
rule out significant errors on both N$^{tar}$ and N$^{proj}$. Even supposing the
concurrence of equally oriented systematic uncertainties on N$_i^{det}$, N$^{tar}$
and N$^{proj}$ it is not possible to explain the magnitude of the discrepancy
and its dependence on $E_{cm}$. The most likely candidate source of error
is then represented by the neutron detection efficiency in the exclusive
measurement \citep{ISH06}.
Indeed, in comparison to $^{11}$B inclusive measurements \citep{BOY92,GU95},
the sensitivity to reaction events in the $^{11}{\rm B}-n$ exclusive
measurements \citep{ISH06} was governed by the neutron counter. In comparison
to neutron inclusive measurements \citep{LAC08} a substantially different
type of neutron detector was used in \citet{ISH06}.
As shown in Fig.\ref{fig2}, the thermalization counter used in \citet{LAC08}
is a zero-energy-threshold detector, its detection efficiency staying at a
significant level down to thermal energies. Instead, the plastic scintillator
array in \citet{ISH06} shows a steep drop in detection efficiency with
decreasing neutron energy $E_{n}$ below 2 MeV with a seeming cut at
$E_{n}=0.5$ MeV. The occurrence of such a detection threshold plays a critical
role in the measurement of the total cross section according to Eq.\ref{eq0}.
With this respect, we consider the calculated kinematical
diagram for the $^{11}$B final states of the ${}^8$Li+${}^4$He$\to{}^{11}$B+$n$
reaction at $E_{cm}=0.75$ MeV, shown in Fig.\ref{fig1bis}. In this plot,
with the cut at 0.5 MeV in the neutron energy spectrum,
the $i=6$ reaction branch is experimentally unaccessible.
In such a situation the exclusive experiment \citep{ISH06} could not provide
the wanted cross sections summed over all $^{11}$B final states.
Moreover, with the information available in \citet{ISH06,Ash06} we have performed
Monte Carlo simulations by implementing the detector set-up in \citet{ISH06,Ash06} into
a GEANT code as described in \citet{CEL97}. Excellent agreement with the
bulk of the experimental efficiency curve is obtained by the dashed histogram in Fig.\ref{fig2}.
This shows a steep drop below 2 MeV, very similar to a sharp detection cut-off,
characterized by the effective half-drop  energy threshold $E_{n}^{cut}\cong 1$ MeV.
With reference to Fig.\ref{fig1bis}, a cut-off of about 1 MeV would make
unaccessible the branches $i=4,5$ besides the $i=6$ reaction branch.
Quantitatively, the more
favored the associated branching ratios the larger is the missing cross section
in the exclusive experiment. It should be noted that the case $E_{cm}=0.75$ MeV
illustrated in Fig.\ref{fig1bis}
is in the hearth of the Gamow window for Big Bang nucleosynthesis.

\begin{figure}
\epsscale{1}
\vspace{-2cm}
\plotone{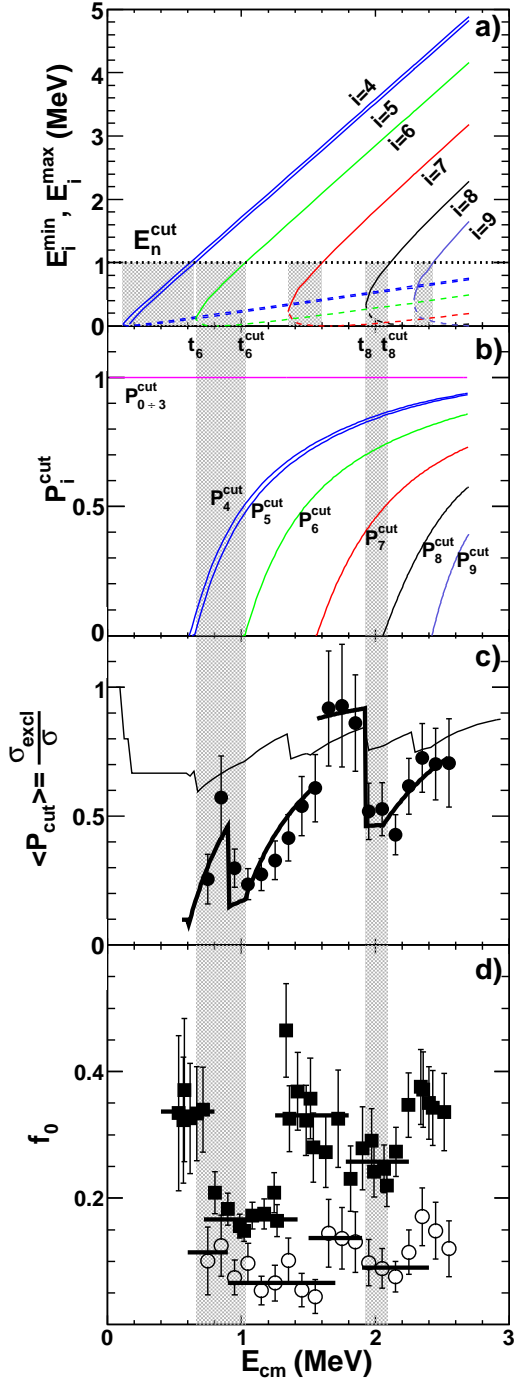}
\caption{\label{fig3} a) Kinematical loci of the minimum (dashed) and
maximum (solid) laboratory neutron energies; b) the branch observability
factors $P_i^{cut}$ for $E_{n}^{cut}=$1 MeV.
The  curves of the branches $i=0\div3$, not shown above, are
entirely above the $E_{n}^{cut}=1$ MeV level and the corresponding $P_{i}^{cut}=1$
independent of $E_{cm}$; (c) the experimental $<P_{cut}>$ (symbols), its best fit
(thick curves) and the uniform pattern case (thin curve); (d) the branching ratio $f_0$
deduced from ground state cross section data in \citet{ISH06} (open circles) and \citet{Para90}
(filled squares x2). The horizontal lines indicate the weighted mean $f_0$ values in the
corresponding intervals.}
\end{figure}

The laboratory reaction
neutron kinematics is summarized in Fig.\ref{fig3}a as a function of $E_{cm}$.
For each reaction branch, at a fixed $E_{cm}$, the
laboratory neutron energy ranges between the minimum $E_{i}^{\min}$ (dashed) and
the maximum $E_{i}^{\max}$ (solid) curves. Assuming $E_{n}^{cut}=1$ MeV, we have
calculated the $E_{cm}$ values $t_i^{cut}$ at which $E_{i}^{\max}$
crosses such $E_{n}^{cut}$ level (dashed black line in Fig.\ref{fig3}a).
These are given in Tab.\ref{tab1}
together with the corresponding set of reaction threshold energies
$t_i$ at which a branch starts to be potentially active, fixed by the corresponding reaction Q-values.
Each $t_i-t_i^{cut}$ couple singles out a $E_{cm}$ kinematic region where
the observation of the corresponding $i-$th branch is completely missed,
because the corresponding laboratory neutron energies stay below the
experimental threshold. These $t_i\div t_i^{cut}$
intervals are emphasized by shaded bands in Fig.\ref{fig3}a. Inside each $E_{cm}$
shaded interval, no efficiency correction can by any means be performed for the
corresponding completely missed branch.

Whenever $E_{cm}$ falls
outside the shaded region of an open branch this becomes
partially observable and its missed portion  is recoverable, only provided
that the observed portion is corrected by the appropriate value of its
observability factor $P_i^{cut}$. For each reaction branch,
this is defined as the observable portion of the laboratory neutron energy
distribution $\frac{dP_i}{dE_n} $ characterized by $E_{n}\ge E_{n}^{cut} $.

Consequently, in the exclusive
measurements \citep{ISH06}, the excitation function, i.e. the cross section versus $E_{cm}$,
can be corrected at some $E_{cm}$-values, but remains necessarily uncorrected at some
other $E_{cm}$ values, right inside each shaded interval,  where the $P_i^{cut}$ of
the corresponding $i=6\div 9$ reaction branch equals zero.

Conversely, the zero-energy detection
threshold inclusive measurements \citep{BOY92,GU95,LAC08} do measure
the wanted cross section $\sigma$ in the whole investigated range of $E_{cm}$.
Accordingly, we write
\begin{equation}
\label{eq1}
\sigma =\sigma\cdot\sum_{i=0}^
{i=i_{\max}} f_i\equiv \sigma_{incl}
\end{equation}
where the factors $f_i$ are the unknown branching ratios as functions of $E_{cm}$.
In Eq.\ref{eq1}, $\sigma_{incl}$ is
the weighted linear interpolation of all inclusive cross section data \citep{LAC08,BOY92,GU95}
within 0.1 MeV bins, represented by the curve in Fig.\ref{fig1}.

On the other hand,
in the case of the exclusive
measurements \citep{ISH06} the detection efficiencies $\epsilon_i$ must incorporate the
observability factor $P_{i}^{cut}$ of the corresponding branch versus $E_{cm}$.
Assuming isotropic neutron emission in
the centre-of-mass system, following the measurements in \citet{ISH06}, we have calculated  the $P_{i}^{cut}$ values for
different $E_{n}^{cut}$. The results for $E_{n}^{cut}= 1$ MeV are displayed in Fig.\ref{fig3}b.
We remark that with the higher
threshold $E_{n}^{cut}\sim 1$ MeV suggested by Fig.\ref{fig2} the portions of missed events
in the exclusive experiment by \citet{ISH06},
and hence the corrections, are much larger than those corresponding to the seeming
cut of 0.5 MeV so that $\sigma_{excl}$ data have remained essentially
uncorrected in \citet{ISH06}. Accordingly, the experimental cross section $\sigma _{excl}$ reported in
\citet{ISH06} can be written as
\begin{eqnarray}
\label{eq2a}
\sigma_{excl} =\sigma \cdot <P_{cut}>\\
\label{eq2b}
<P_{cut}>=\sum_{i=0}^{i=i_{\max}}
f_{i} \cdot P_{i}^{cut}
\end{eqnarray}
$<P_{cut}>$ being the average of the observability factors $P_i^{cut}$
weighted by the branching ratios. Eqs.\ref{eq2a}-\ref{eq2b} not only clarify the primary
role played by the experimental threshold $E_n^{cut}$ through the factors
$P_{i}^{cut}$ $(\leq 1)$, but also the
fundamental concurrent role played by branching ratios. Indeed, we underscore that
if all $P_{i}^{cut}\cong 1$, $\sigma_{excl}\cong \sigma$
no matter the branching ratio pattern because $\sum_{i=0}^{i=i_{\max}}f_i=1$
and, consequently, $<P_{cut}>\cong 1$.
Conversely, if at least one of the $P_{i}^{cut}$ is significantly
smaller than one $\sigma_{excl}$ underestimates $\sigma$,  the more
favored the feeding of a (partially) missed branch, the larger the deviation of
$\sigma_{excl}$ from $\sigma$. With reference to the case in Fig.\ref{fig1bis},
this effect is more significant when reaction branches involving large $^{11}$B excitation energies
come into play, their observation being extremely sensitive to the neutron threshold energy $E_n^{cut}$.
Clearly, the deviations of $\sigma_{excl}$ from $\sigma_{incl}$ are entirely
described  by $<P_{cut}>$. Accordingly, from   Eq.\ref{eq2a}, we have
deduced the experimental $<P_{cut}>$
of the exclusive measurement \citep{ISH06} as $<P_{cut}>=\frac{\sigma _{excl}}{\sigma_{incl} }$,
the errors of the interpolated $\sigma_{incl}$ being appropriately propagated.
It is shown in Fig.\ref{fig3}c versus $E_{cm}$
and confirms that the largest deviations between $\sigma_{excl}$ and $\sigma_{incl}$
occur right in correspondence of the shaded bands, forming a
marked saw-tooth-like behavior with two apparent falls corresponding to the opening
of the $i=6$ and $i=8$ branches.

To describe such a rise-and-fall behavior, a simple recurrence formula
can be used, following immediately from Eq.\ref{eq2b}:
\begin{equation}
\label{eq3}
<P_{cut}^{(+)}>=(1-f_{{i_{max}}})\cdot <P_{cut}^{(-)}>+f_{i_{max}}\cdot P_{i_{max}}^{cut} \, ,
\end{equation}
where $(-)$ and $(+)$ denote near-threshold $E_{cm}$ values on the left and right side of
$t_{i_{max}}$, respectively. The two terms on the right side govern the magnitude
of each fall and of the following rise, respectively.

As a reference case,  in Fig.\ref{fig3}c we show the  $<P_{cut}>$ calculated according
to Eq.\ref{eq2b} for the $f_{0\div i_{max}}=(i_{max}+1)^{-1}$ uniform branch pattern
(thin solid line). The energetically open $f_i$'s are assumed constant in each $t_i\div t_{i+1}$
$E_{cm}$ interval and the $P_i^{cut}$ in Fig.\ref{fig3}b, evaluated for $E_n^{cut}=1$ MeV,
are used in the calculation. In this case, falls of $<P_{cut}>$
occur at the successive opening of each of the $i=4\div 9$ reaction branches because the
threshold $E_n^{cut}$ on the neutron energy makes each of them experimentally unaccessible
inside the corresponding $t_i\div t_i^{cut}$ interval.
Strikingly, not only the falls in the experimental $<P_{cut}>$ recall this type
of discontinuity but, in addition, we can conclude that in the experiment the
$i=6$ and $i=8$ branches strongly deviate from an uniform pattern.

Because of the normalization to 1, when the $i=6$ and $i=8$ branches strongly add up,
correspondingly strong falls should be observable by examining the population of
all the lower-i active branches.
Indeed, this is what we have found in the trend of the experimental
$f_0=\frac{\sigma _0}{\sigma_{incl} }$ values as a function of $E_{cm}$, as it is clearly demonstrated in Fig.\ref{fig3}d.
This has been determined here starting both from the
exclusive $\sigma _{0}$ data \citep{ISH06} and from the $\sigma _0$ deduced from
the $^{11}B(n,\alpha)^{8}Li$ inverse reaction in \citet{Para90}.
We also note that an apparent increase of $f_0$ in both  data sets distinguishes two regions, below and above $E_{cm}\sim 1.6$ MeV, where different branch pattern regimes come presumably into play.

Accordingly, two separate  fits of the experimental
$<P_{cut}>$, for the region below and above  $E_{cm}\sim 1.6$ MeV,
has been performed to determine the branching ratios $f_i$, using the subroutine MINUIT.
Following Fig.\ref{fig3}d, the branching ratios $f_{0\div i_{max}}$ are assumed constant inside
each $E_{cm}$ interval in between two successive branch openings.
Therefore, all $f_i$
can be treated as free fitting parameters only constrained by the normalization
condition $\sum_{i=0}^{i_{max}} f_i=1$. In particular, as the factors
$P_{0\div 3}^{cut}=1$ (Fig.\ref{fig3}b), Eq.\ref{eq2b} becomes
$<P_{cut}>\equiv \sum _{i=0}^3f_{i} + \sum _{i=4}^{i_{max}}f_{i} \cdot
P_{i}^{cut}$ so that we have considered the $i=0\div 3$ as a single branch.
Because the threshold energies $t_4$ and $t_5$ are so close (Tab.\ref{tab1}),
also the $i=4\div 5$ are treated as a single branch.
The resulting branching ratio (and error) values are
listed in Tab.\ref{tab1} and the associated curves are shown in Fig.\ref{fig3}c\footnote{The
sensitivity to each $f_i$ is the higher the steeper is the rise of the corresponding
$P_i^{cut}$. This implies that in the $E_{cm}$ region, where the $P_{4-6}^{cut}$
slowly approach unity, the fitting procedure presumably adds small $f_{4-6}$ values to
$f_{0\div 3}$.}. We also remark that both start energy and shape of the two rises in the
experimental $<P_{cut}>$ are extremely sensitive to $E_{n}^{cut}$. By tests performed, this
fact leads to the  independent determination of $E_{n}^{cut}\sim 1$ MeV, in agreement with
the analysis of efficiency data in  Fig.\ref{fig2}.
To cross check these results we evaluate the
$<f_i^{cut}>$ averaged over the $0.75\leq E_{cm}\leq 2.55$ MeV range,
and compare with those given in
\citet{ISH06} (Tab.\ref{tab1}). These $<f_i^{cut}>$ are
linked with the true
$f_i$ by
\begin{equation}\label{grandimedie}
<f_i^{cut}>=\frac{\int_{0.75}^{2.55}\sigma\cdot P_i^{cut}\cdot f_i\cdot
dE_{cm}}{\sum_i\int_{0.75}^{2.55}\sigma \cdot P_i^{cut}\cdot f_i\cdot dE_{cm}}.
\end{equation}
Inserting  into
Eq.\ref{grandimedie} the $f_i$ values established above results in the $<f_i^{cut}>$
reported in Tab.\ref{tab1},
the errors of $\sigma$ and of all the $f_i$ being propagated accordingly. The good agreement
of this comparison strongly supports  the validity of Eq.\ref{eq1}
and, therefore, the reliability  of
$\sigma_{incl}$.


The physical novelty of the present Letter is the apparent
selective feeding of the highest excited $^{11}$B levels, in particular of the $i=6$ and $i=8$
branches leading to the $^{11}$B levels at 7.29 and 8.56 MeV, respectively, originating
the discussed non-uniformities. This peculiar trend can hardly be understood invoking
selection rules only, as there does not seem to be anything unique in terms of quantum
numbers about the $i=6$ and $i=8$ branches (see Table \ref{tab1}). Rather it seems to
signal that the nuclear structure of the initial $^{12}$B* and of the final $^{11}$B*
excited states plays the most important role in determining the characteristic non-uniformity of the
observed branching ratios. Concerning $^{12}$B, in the excitation energy region explored
here ($E^*=E_{cm}+10.01$ MeV), states with large $\alpha$, $t$ or $^5$He spectroscopic factors
have recently been emphasized in the ${}^9{\rm Be}+ {}^7{\rm Li} \to 2\alpha + {}^8{\rm Li}$ reaction.
In particular, clear evidence exists for at least two states, at 10.9 and 11.6 MeV, which  show
significant  $\alpha$  widths \citep{Soic}.
The relative contributions of these $^{12}$B states
to the $^{8}$Li($\alpha$ ,n)$^{11}$B reaction process are regulated by both their $\alpha$ and
$n$ partial widths. The $^{12}$B state at 10.9 MeV clearly contributes, giving rise to a
resonance at $E_{cm}\sim 0.9$ MeV,
as it is demonstrated by $\sigma_{incl}$ (Fig.\ref{fig1}).
The contribution of the 11.6-MeV $^{12}$B state  cannot be excluded as a hump
does appear right at $E_{cm}\sim1.6$ MeV in the cross section $\sigma_{incl}$,
indicating a small $n$-width and, therefore, a possibly complex cluster structure.
Concerning $^{11}$B, a well developed 2$\alpha$+$t$ cluster structure of the $i=8$
state at 8.56 MeV has  been established very recently, whereas the other lower-lying
negative-parity $^{11}$B levels are successfully described by shell-model calculations \citep{Kawa07}.
Right below $t_8$, the trend of the experimental $<P^{cut}>$ is consistent with
the one obtained for the uniform branch pattern, which suggest an at most weakly non-uniform preferential
feeding of the lowest energy $i=0\div 3$ reaction branches. This can be likely attributed to a mismatch
between initial- (cluster) and final- (single particle) state nuclear structures.
Above $t_8$, the enhanced relative feeding of the newly open $i=8$ branch,
signalling a large overlap between the initial and final state wave functions,
indicates a  $^{12}$B cluster structure close to the 2$\alpha$+$t$ one of the 8.56 MeV level
of the daughter $^{11}$B nucleus (plus neutron).

The comparable reaction branch non-uniformity
type established here for the $i=6$ branch suggests close structures of both  $^{11}$B and
$^{12}$B involved states.  The relevant issues pointed out here call for further investigations.

In conclusion, in this Letter we have shown robust evidences that support the
large values from inclusive measurements \citep{LAC08,GU95,BOY92} as representing
at present the most reliable estimate of the ${}^8$Li+${}^4$He$\to{}^{11}$B+$n$
reaction cross section at astrophysical energy.
The recommended value of this cross section (and the corresponding error $\Delta\sigma$),
following the considerations developed above, is:
\begin{eqnarray}\label{recc_CS}
\sigma(E_{cm})=\sum_{k=0}^1a_kE_{cm}^k+\sum_{j=1}^3\frac{b_j}{(E_{cm}-E_j)^2+g_j^2/4}\\
\Delta\sigma=\sum_{n=0}^2c_nE_{cm}^n
\end{eqnarray}
where $a_0=-9.09$~mb, $a_1=97.2$~mb/MeV, $b_1=18.6$~mb\,MeV$^2$,
$b_2=-514$~mb\,MeV$^2$, $b_3=722$~mb\,MeV$^2$, $g_1=0.509$~MeV,
$g_2=0.914$~MeV, $g_3=1.05$~MeV, $E_1=0.986$~MeV, $E_2=1.84$~MeV, $E_3=1.87$~MeV,
$c_0=66.9$~mb, $c_1=-18.7$~mb/MeV, $c_2=11.5$~mb/MeV$^2$.
The previous formula is a simple fitting of the cross-section weighted linear interpolation
in Fig.\ref{fig1}, whose accuracy is better than 5\% in the whole energy range.

The original approach we have
developed here can have important applications in different fields and can be extended to
become an effective
experimental method to extract spectroscopic information otherwise inaccessible with
present-day experimental facilities.
For the nuclear physics case considered here we have determined for the first time
significantly non-uniform branch patterns, which are interpreted as manifesting the
exotic cluster structure recently discovered in $^{11}$B \citep{Kawa07} and  $^{12}$B
\citep{Soic} excited nuclei. More importantly, this Letter strongly calls for revised
calculations of the r-process nucleosynthesis. Indeed, \citet{Sasa06} concluded that
the entropy per baryon increases by
about a factor of 2 from previous estimates in \citet{Sasa05}, using the cross section in \citet{ISH06}.
According to the present revised cross-section value  of the
${}^8$Li+${}^4$He$\to{}^{11}$B+$n$ reaction, the consequent constraint on
models of the r-process astrophysical site might be significantly
altered, with undoubtedly interesting consequences for astrophysics.
We will explore possible additional implications of this
work in future studies.

\clearpage

\clearpage

\clearpage

\clearpage

\clearpage

\begin{landscape}
\begin{table}
\begin{center}
\caption{\label{tab1}$^{11}$B final state, center-of-mass reaction energy
threshold $t$, observability energy threshold $t^{cut}$ (assuming $E_{n}^{cut} = 1$ MeV),
true branching ratios $f$ in the indicated $E_{cm}$ intervals and
mean branching ratios $<f^{cut}>$. Short lines indicate
branching ratio values compatible with zero. Empty entries
correspond to branches which are not physically open.
In the fitting procedure, the $i=0\div 3$ branches are not
separated and the summed $f_{0\div 3}$ is given. The same applies to the $i=4,5$ branches.}
\begin{tabular}{crrrrrrrrr}
\tableline\tableline
i & State in $^{11}$B & $t_i$ & $t_i^{cut}$ & \multicolumn{4}{c}{$f_i$} & \multicolumn{2}{c}{$<f_i^{cut}>$}\\
& J$^{\pi}$, E$^{*}$ (MeV) & (MeV) & (MeV) & $0.75\div 0.9$ MeV & $0.9\div 1.6$ MeV & $1.6\div 1.9$ MeV &
$1.9\div 2.55$ MeV & this work & Ref. \citep{ISH06} \\
\tableline
0 & $\frac{3}{2}^{-} \quad 0.00 $ & $0$  & $0$  & \multirow{4}{*}{$0.16\pm 0.11$} & \multirow{4}{*}{0.05$\pm 0.03$}
& \multirow{4}{*}{0.87$\pm 0.16$} & \multirow{4}{*}{0.43$\pm$ 0.08} & \multirow{4}{*}{$0.62\pm 0.05$} & \multirow{4}{*}{$0.547\pm 0.042$} \\
1 & $\frac{1}{2}^{-} \quad 2.12 $ & $0$  & $0$  & & & & & & \\
2 & $\frac{5}{2}^{-} \quad 4.44 $ & $0$  & $0$  & & & & & & \\
3 & $\frac{3}{2}^{-} \quad 5.02 $ & $0$  & $0$  & & & & & & \\
4 & $\frac{7}{2}^{-} \quad 6.74 $ & 0.11 & 0.61 & \multirow{2}{*}{$0.84\pm 0.11$} & \multirow{2}{*}{$0.27\pm 0.11$} & \multirow{2}{*}{$-$}
  & \multirow{2}{*}{$-$} & \multirow{2}{*}{$0.14\pm 0.03$} & \multirow{2}{*}{$0.181\pm 0.027$}\\
5 & $\frac{1}{2}^{+} \quad 6.79 $ & 0.16 & 0.65 &             &            &             &  &  &  \\
6 & $\frac{5}{2}^{+} \quad 7.29 $ & 0.66 & 1.03 & $-$            & $0.68\pm 0.10$ & $-$            & $-$ & $0.12\pm 0.02$ & $0.132\pm 0.015$ \\
7 & $\frac{3}{2}^{+} \quad 7.98 $ & 1.35 & 1.60 &             & $-$            & $0.13\pm 0.16$ & $0.06\pm 0.09$ & $0.04\pm 0.04$ &
$0.117\pm 0.018$ \\
8 & $\frac{3}{2}^{-} \quad 8.56 $ & 1.93 & 2.11 &             &             &             & $0.50\pm 0.08$ & $0.08\pm 0.02$ &
$0.024\pm 0.005$ \\
\tableline
\end{tabular}
\end{center}
\end{table}
\end{landscape}


\begin{thebibliography}{}
\bibitem[Boyd et al.(1992)]{BOY92} R.N. Boyd et al., 1992, Phys. Rev. Lett. 68, 1283
\bibitem[Celano et  al.(1997)]{CEL97} L. Celano et  al., 1997, Nucl. Instr. Meth. A 392, 304
\bibitem[Gu et al.(1995)]{GU95}  X. Gu et al., 1995, Phys. Lett. B 343, 31
\bibitem[Hashimoto et al.(2006)]{Ash06} T. Hashimoto et al., 2006, Nucl. Instr. Meth. A 556, 339
\bibitem[Ishiyama et al.(2006)]{ISH06} H. Ishiyama et al., 2006, Phys. Lett. B 640, 82
\bibitem[Kajino \& Boyd(1990)]{KAJ90} T. Kajino \& R. Boyd, 1990, Astrophys. J. 359, 267
\bibitem[Kawabata et al.(2007)]{Kawa07} T. Kawabata et al., 2007, Phys. Lett. B 646, 6
\bibitem[La Cognata et al.(2008)]{LAC08} M. La Cognata et al., 2008, Phys. Lett. B. 664, 157
\bibitem[Lara et al.(2006)]{Lara06} J.F. Lara et al., 2006, Phys. Rev. D 73, 083501
\bibitem[Malaney \& Fowler(1988)]{MAL88} R.A. Malaney \& W.A. Fowler, 1988, Astrophys. J. 333, 14
\bibitem[Mao et al.(1994)]{Mao2} Z.Q. Mao et al., 1994, Nucl. Phys. A 567, 125
\bibitem[Matsuura et al.(2005)]{MAT05} S. Matsuura et al., 2005, Phys. Rev. D 72, 123505
\bibitem[Paradellis et al.(1990)]{Para90} T. Paradellis et al., 1990, Z. Phys. A 337, 211
\bibitem[Rauscher et al.(2007)]{RAU07} T. Rauscher et al., 2007, Phys. Rev. D 75, 068301
\bibitem[Sasaqui et al.(2005)]{Sasa05} T. Sasaqui et al., 2005, Astrophys. J. 634, 534
\bibitem[Sasaqui et al.(2006)]{Sasa06} T. Sasaqui et al., 2006, Astrophys. J. 645, 1345
\bibitem[Soic et al.(2003)]{Soic} N. Soic et al., 2003, Europhys. Lett. 63, 524
\bibitem[Terasawa et al.(2001)]{TER01} M. Terasawa et al., 2001, Astrophys. J. 562, 470
\end{thebibliography}
\end{document}